\documentstyle[12pt]{article}

\begin{document}

\centerline{\large \bf  On the lagrangian of N=1, D=10 dual supergravity }
\bigskip
\centerline{{\bf  K.N.Zyablyuk}\footnote{The research described in this
publication was made possible in part by the grant no. MOY000 from the
International Science Foundation.}}
\bigskip
\centerline{\it Institute of Theoretical and Experimental Physics,}
\centerline{\it Moscow State University}

\bigskip

\begin{abstract}
It is shown how one can construct the lagrangian of dual supergravity
by means of the equations of motion derived from the superspace approach.
\end{abstract}

Dual N=1, D=10 supergravity is equivalent to the usual N=1, D=10
supergravity at the level of the minimal lagrangian (i.e. one
which contains not more than 2-nd derivatives). But the condition of
anomalies cancellation implies that some nonminimal terms
must be added to the lagrangian. The anomalies can be taken
into account in the superspace approach. It arrears (see \cite{3,5}),
that in the usual supergravity the nonminimal corrections turn
out to be the infinite series of terms each of them should be
calculated by the perturbation theory in the string coupling
constant. Meanwhile the dual
supergravity lagrangian has a finite number of terms which can
be written explicitely \cite{2}. But the superspace approach gives
us only the equations of motion; the building of the lagrangian
is a matter of some difficulty. Initially the dual
supergravity lagrangian has been obtained in \cite{6} by means
of the dual transformation from the usual supergravity lagrangian.
Some partial results on the nonminimal corrections
have been obtained in \cite{5,1} but even the bosonic part of
the lagrangian has not been found completely. The main difficulty is
to transform the equations of motions to the form where they
immediately follow from the lagrangian. The presence of additional
constraints (see below) also complicates the situation.Our
purpose here is to formulate a method, which helps to resolve the
difficulties. In this paper the
minimal lagrangian of N=1, D=10 dual supergravity is constructed
by means of the equations of motion, derived from the superspace
approach. We hope that this work helps us to build the anomaly
free lagrangian when the nonminimal corrections
to the equations of motion will be written explicitely.

The multiplet of the dual supergravity contains:
 ${e_m}^a$-- graviton, ${\psi_m}^\alpha$-- gravitino,
 $\phi$ -- dilaton, $\chi_\alpha$-- dilatino and $M_{m_1 \ldots m_6}$--
antisymmetrical potential (we do not consider the contribution of
the gauge matter here). The equations of motion for these fields
depend on the constraints imposed on the components of
the correspondent superfields. We use the constraints introduced
in \cite{8} which are supposed to be the simplest ones. Equations of motion
in this system of constraints (after some fields redefinition)
 have been obtained in \cite{2} and are listed in Appendix A of
this paper. These equations have the important property: they are
linear in fields $\phi$ and $\chi$. Hence the lagrangian must be
linear in dilaton and dilatino fields:

$$ L = (\,\phi\,S_1 + \chi_\alpha\,S_2^\alpha\,) \vert        \eqno(1) $$
(the vertical line denotes the 0-component of a superfield).

In order to find $S_1$ and $S_2^\alpha$ it is necessary to use the
torsion Bianchi identities. In this set of constraints they do not fix
the curvature ${\cal R}_{abcd}$ and the torsion $T_{abc}$ and
${T_{ab}}^\alpha$ components. The general expressions for $S_1$ and
$S_2^\alpha$ through these components, which contain not more than
the 2-nd derivatives, are:

$$ S_1 = {\cal R} + a\,T^2 ,\;\; S_2 = b\,\Gamma^{ab}T_{ab} ,  \eqno(2) $$
where ${\cal R}={{\cal R}_{ab}}^{ab}$ , $T^2=T_{abc}T^{abc}$ ,
the $T_{ab}$ denotes ${T_{ab}}^\alpha$; the common numerical
factor before the lagrangian is not
important here. The quantities $S_1$ and $S_2$ must vanish on shell.
 But the torsion Bianchi identities guarantee, that the following
relations are valid on shell (see \cite{2})

$$ {\cal R} - {1\over 3}\,T^2 = 0                         \eqno(3) $$

$$ \Gamma^{ab}T_{ab} = 0                                \eqno(4) $$
(in the set of constraints presented in Appendix A; however the
analogous relations take place in some other systems of constraints,
 for instanse in \cite{3,4}). Hence $a=-1/3$.

In order to find factor $b$ it is necessary to consider some
equation for ${e_m}^a$, $\psi_m$ or $M_{m_1 \ldots m_6}$. For this
purpose we must write down the 0-components (see Appendix B) of
the relations (A.3) -- (A.7) and transform them to the form where
they may be derived from the lagrangian immediately.

We demonstrate, how one can do it for the potential $M_{m_1 \ldots m_6}$.
 To write down the 0-component of the relation (A.5) let us
express the derivative with the flat index $D_a$ through the derivative
with the world index $D_m$:

$$ D_a\vert = {e_a}^m D_m \vert - {1\over 2}\,{\psi_a}^\alpha
 D_\alpha \vert                                     \eqno(5) $$
We take the spinor derivative $D_\alpha T_{abc}$ from the solution of
the torsion Bianchi identities (see \cite{2}):

$$ DT_{abc} = -{1\over 2}\,{\Gamma_{abc}}^{fh} T_{fh} +
 \alpha\,\Gamma_{abc}\Gamma^{fh} T_{fh}             \eqno(6) $$
(the Bianchi identities give us only on-shell fields values, i.e. while
 $\Gamma^{fh}T_{fh}=0$; so one can choose arbitrary $\alpha$ in (6)).
Then let us substitute the expressions (B.3) and (B.4) for the connection
 ${\phi_{ma}}^b\vert$ into the derivative $D_m$ defined according
to (B.2) and after that substitute the torsion components
${T_{ab}}^\alpha\vert$ and $T_{abc}\vert$ from (B.5) and (B.6).

The terms

$$ \beta\,\psi_{[a}\Gamma_{bc} (\,\phi\,\Gamma^fT_{d]f} - D_{d]}\,\chi -
{1\over 36}\,\Gamma_{d]} {\hat T}\,\chi - {1\over 24}\,{\hat T}\Gamma_{d]}
\,\chi\,)\vert = 0                      \eqno(7) $$

$$ \gamma\,\psi_{[a}\Gamma_{bcd]} (\, {\hat D}\,\chi +
{1\over 9}\,{\hat T}\,\chi\,)\vert = 0  \eqno(8) $$
are equal zero on shell and thus can be added to resulting
expression with arbitrary factors
 $\beta$ and $\gamma$. However the lagrangian must be invariant
relative to the transformation
$$ M_{m_1 \ldots m_6} \rightarrow M_{m_1 \ldots m_6} +
\partial_{[m_1} f_{m_2 \ldots m_6]}\;,       \eqno(9)          $$
which does not involve other fields. Consequently it must contain
$M_{m_1 \ldots m_6}$ only throught the field-strenght
$\partial_{[m_1} M_{m_2 \ldots m_7]}$. So the variation of action with
respect to the potential must be the full derivative. This claim fixes all
unknown factors unambiguously: $\alpha = 1/2$, $\beta= 3/2$, $\gamma= 0$.

As a result we derive the equation of motion for $M_{m_1 \ldots m_6}$:

$$ (\,\phi\,M_{[abc} - {1\over 8}\,\phi\,\psi_f\Gamma^{[f}\Gamma_{[abc}
\Gamma^{h]}\psi_{\vert h \vert} + {3\over 2}\,\psi_{[a}\Gamma_{bc}\,
\chi\,)_{;\,d]} = 0                                        \eqno(10) $$
where $$ M_{abc} \equiv {1\over 6!}\,{\epsilon_{abc}}^{d_1\ldots d_7}
M_{d_1\ldots d_6;\,d_7}\;\;,  $$
the semicolon denotes the ordinary covariant derivative which
depends on the ordinary vielbein.

Taking into account (see (B.5), (B.6), (B.8)) that 0-components are
$$ {\cal R}\vert = {1\over 4}\,M_{abc}^2 + \ldots;\;\;
T_{abc}\vert = M_{abc} + \ldots;\;\;
\Gamma^{ab}T_{ab}\vert = -{1\over 8}\,M_{abc}\Gamma^{ab}\psi^c + \ldots; $$
where the dots denote the terms of the lower order in the M-field,
 it is easy to see that the equation (10) can be derived from the lagrangian
 (1),(2) with $b=2$.

So the lagrangian must be the 0-component of the following stuff:

$$ {\cal L} = \phi\,(\,{\cal R} - {1\over 3}\,T^2\,) +
 2\,\chi\,\Gamma^{ab}T_{ab}                                      \eqno(11) $$

But there is an ambiguity in the field representation of the quantity
 ${\cal R}\vert$ at the off-shell level.
Indeed, it is necessary to know the curvature
components with the spinor indices in order to calculate ${\cal R}\vert$
 from the formula ${\cal R}\vert={E_b}^N{E_a}^M{{\cal R}_{MN}}^{ab}\vert$.
 We can take it from the solution of the torsion Bianchi identities in
\cite{2}. But as mentioned before the Bianchi identities give us only
on-shell fields values. Consequently ${\cal R}\vert$ may be presented in
different forms which are equivalent up to the term

$$ \psi_a\Gamma^a\Gamma^{bc} T_{bc} \vert\;\;,           \eqno(12) $$
vanishing  on-shell. One of the possible variants is presented in
Appendix B eq. (B.8). It contains terms with $M_{abc}$ which lead to
$$ (\,{\cal R} - {1\over 3}\,T^2\,)\vert = -{1\over 12}\,M_{abc}^2 +
{1\over 48}\,\psi_a\Gamma^{[a}{\hat M}\Gamma^{b]}\psi_b +\ldots   $$
where ${\hat M} = M^{abc}\Gamma_{abc}$.

Substituting this expression into (11) one can derive the lagrangian
which leads to the true equation for $M_{m_1 \ldots m_6}$ (10). But
if we added the term (12) to (B.8) we would get a wrong equation,
 different from (10). So the only expression for ${\cal R}\vert$
 suitable for us is the eq. (B.8).

We see that equations (3), (4), (10) fix all the terms in the lagrangian
without the equations of motion for ${e_m}^a$ and ${\psi_m}^\alpha$.
 But the straightforward calculation demonstrates that variations of $L$ with
respect to ${e_m}^a$, ${\psi_m}^\alpha$-- fields vanish if one takes into
account the equations of motion (A.3) -- (A.7), derived from the
solution of Bianchi identities. So all the assumptions about the
lagrangian structure are confirmed.

Finally:

$$ L = \phi\,R  - {1\over 2}\,\phi\,\psi_a\Gamma^{abc}\psi_{c;\,b} -
{1\over 2}\,\phi_{;\,a}\psi^a\Gamma_b\psi^b +
2\,\psi_a\Gamma^{ab}\chi_{;\,b} +                                           $$
$$- {1\over 12}\,\phi\,M_{abc}^2 + {1\over 48}\,\phi\,\psi_a\Gamma^{[a}
{\hat M}\Gamma^{b]}\psi_b - {1\over 4}\,\chi\,\Gamma^{ab}\psi^c M_{abc} -   $$
$$ - {1\over 3\cdot 256}\,\phi\,{(\psi^d\Gamma_{dabcf}\psi^f)}^2 +
{1\over 64}\,\phi\,{(\psi_a\Gamma_b\psi_c)}^2 +
{1\over 32}\,\phi\,(\psi^a\Gamma^b\psi^c)(\psi_a\Gamma_c\psi_b) -           $$
$$ - {1\over 16}\,\phi\,{(\psi_a\Gamma^b\psi_b)}^2 +
{1\over 8}\,(\chi\Gamma_{ab}\psi_c)(\psi^a\Gamma^c\psi^b) -
{1\over 4}\,(\chi\Gamma_a\Gamma_b\psi^b)(\psi^a\Gamma_c\psi^c)    \eqno(13) $$

We shall write below the supersymmetry transformations for the fields of the
gravity multiplet. They are the variations of the correspondent superfields
under the shift $\delta z^M=\varepsilon^M$, $\varepsilon^M\vert=
(0,\varepsilon^\alpha)$ and may be easily obtained for
 ${e_m}^a, \psi_m, \phi$ and $\chi$. To derive the supersymmetry
transformation for the $M_{m_1 \ldots m_6}$ it needs to write the
0-component of variation $\delta T_{abc}=\varepsilon^\alpha D_\alpha T_{abc}$,
taking $D_\alpha T_{abc}$ from (6) and $T_{abc}\vert$ from (B.6).
 Choosing $\alpha=0$ in (6) one can transform this variation to the form:
$$ {\epsilon_{n_1 n_2 n_3}}^{m_1 \ldots m_7}(\,\delta M_{m_1 \ldots m_6}
+ 3\,\psi_{m_1}\Gamma_{m_2 \ldots m_6}\varepsilon\,)_{;\,m_7} = 0 \eqno(14) $$
The relation (14) defines $\delta M_{m_1 \ldots m_6}$ unambiguously
up to the transformation of the form (9).

So the supersymmetry transformations in the system of constraints (A.1)
 and (A.2) are:

$$ \delta {e_m}^a = {1\over 2}\,\psi_m\Gamma^a\varepsilon     $$
$$ \delta \psi_m = 2\,\varepsilon_{;\,m} - {1\over 72}\,
(\,3{\hat T}\Gamma_m + \Gamma_m{\hat T}\,)\vert\,\varepsilon -
{1\over 2}\,C_{mpq}\Gamma^{pq}\varepsilon                     $$
$$ \delta \phi = - \chi\,\varepsilon                            $$
$$ \delta \chi = - {1\over 2}\,\phi_{;\,m}\Gamma^m\varepsilon +
{1\over 4}\,(\psi_m\,\chi)\Gamma^m\varepsilon +
{1\over 36}\,\phi{\hat T}\vert\,\varepsilon                     $$
$$ \delta M_{m_1 \ldots m_6} = - 3\,\psi_{[m_1}
\Gamma_{m_2 \ldots m_6]}\,\varepsilon                 \eqno(15) $$
where $$ C_{klm}={1\over 8}\,(\,2\,\psi_k\Gamma_{[l}\psi_{m]} +
\psi_l\Gamma_k\psi_m\,)\;\;.  $$

A lot of work need to be done to check the invariance of lagrangian
 (13) relative to transformations (15). The following speculations
help to reduce it.

Let us divide the supersymmetry transformations by two parts:
$$ \delta\int eL = \int e\,(A + B)\,\varepsilon   \eqno(16)  $$
where $e=\det {e_m}^a$. The part $A$ contains all the terms which can
be written symbolicaly in powers of fields and derivatives:
$$ A = (\chi + \phi\psi)\times(R + \partial^2 + M\partial +
\psi^2\partial)                                          $$
Terms in $B$ do not contain derivatives:
$$ B = (\chi + \phi\psi)\times(M^2 + \psi^2 M + \psi^4)  $$
(except of the derivative inside $M$: here $M$ denotes
the field-strenth $M_{abc}$ but not the potential).

Straightforward calculations show that $A=0$. Then it appears that $B=0$ too.

To prove this let us suppose that fields obey the equations of motion.
 Then one can express the terms of kind $A$ through the terms
of kind $B$ and substitute them in the right-hand side of (16).
 Note, that the left-hand side of (16) must vanish on-shell then
 $B=-A=0$. But the term $B$ does not contain derivatives, therefore
it is not changed after this substitution. So $B=0$ in any case,
 not only on the mass-shell.

Let us suppose now that some terms with derivatives remain in $A$
 because they cannot be expressed through the terms without derivatives
by means of the equations of motion. In this case we have a new
differential equation $A=0$. But all the field equations
are the equations of motion or their consequences. As a result
we came to the contradiction.

So in order to check the invariance of action relative to the
supersymmetry transformations it is necessary to show only that $A=0$.
 It was made by the author of this paper.

The lagrangian (13) does not contain the kinetic terms of dilaton
and dilatino.\footnote{Absence of the dilaton kinetic term was noted in
 \cite{5} for the lagrangian obtained there.} However, it is possible to
transform the Einstein's term of action $\phi R$ to the canonical form
by means of the vielbein conformal transformation. Dilaton kinetic
term appears as a result. The dilatino kinetic term could arise if one
diagonalize the terms with the derivatives of $\psi_m$ and $\chi$ in (13).
 Indeed, the field change $\{\Phi_i\}\rightarrow\{{\Phi_i}'\}$

$$ {e_m}^a = e^{\phi'/6}\,{{e_m}^a}'                                      $$
$$ \psi_m = 2\,e^{\phi'/12}\,(\,{\psi_m}' -
{1\over 6\sqrt{2}}\,{\Gamma_m}'\chi'\,)                                   $$
$$ \phi = e^{-4\phi'/3}                                                   $$
$$ \chi = -{2\sqrt{2}\over 3}\,e^{-17\phi'/12}\,\chi'                     $$
$$ M_{m_1 \ldots m_6} = 2\,{M_{m_1 \ldots m_6}}'                \eqno(17) $$
leads to the lagrangian $L'$
$$ \int e\,L = 4\int e'\,L'  $$
with the canonical kinetic terms:
$$ L' = {1\over 4}\,R' + {1\over 2}\,{\phi^{;\,a}}'{\phi_{;\,a}}' -
{1\over 12}\,e^{-2\phi'}\,{M_{abc}^2}' -
{1\over 2}\,{\psi_a}'\Gamma^{abc}{\psi_{c;\,b}}' +
{1\over 2}\,\chi'\Gamma^a{\chi_{;\,a}}' -                               $$
$$ -{1\over \sqrt{2}}\,{\phi_{;\,b}}'{\psi_a}'\Gamma^b\Gamma^a\,\chi' +
{1\over 24}\,e^{-\phi'}\,(\,{\psi_a}'\Gamma^{[a}{\hat M}'\Gamma^{b]}
{\psi_b}' + {\sqrt{2}}\,\chi'\,\Gamma^a{\hat M}'{\psi_a}'\,)+
\ldots \;,                                                    \eqno(18) $$
where the dots denote the terms of 4-th order in fermionic fields.
 It is easy to see that the lagrangian (18) is the lagrangian
of dual supergravity \cite{6} rewritten in the notations used here.

Note in conclusion that the lagrangian (13) in fields parametrization
 \cite{2} is simpler than lagrangian \cite{6} in fields parametrization
 \cite{7}. It allows us to hope that nonminimal terms which must be
added for cancellation of anomalies are more simple in this
parametrization too. In particular, they must be independent of the
$\phi$ and $\chi$ fields, because variations of the action with respect
to these fields (A6) and (A7) have no anomalous corrections \cite{2}.

The author thanks M.V.Terentjev for the formulation of the problem
and the help in the solution of it.

\bigskip

{\bf Appendix A: constraints and equations of motion}

\bigskip

The constraints, the solution of Bianchi identities and the equations
of motions are taken from \cite{2}. ${\cal R}_{abcd}$ here
corresponds to $-R_{abcd}$ in \cite{2}; all other
notations are the same. All necessary properties of $\Gamma$-matrices
one can find in \cite{7}.

We use the following constraints for the torsion components:
$$ {T_{\alpha\beta}}^c = {\Gamma_{\alpha\beta}}^c,\;\;\;
{T_{\alpha\beta}}^\gamma = {T_{\alpha b}}^c =0    $$
$$ {T_{a\beta}}^\gamma = {1\over 72}\,{{({\hat T}\Gamma_a)}_\beta}^\gamma,
\;\;\;\mbox{where}\;\;\;{\hat T} = T^{abc}\Gamma_{abc}\,.      \eqno(A.1) $$
The Bianchi identities for the 7-form $N=dM$ relate the torsion components
 $T_{abc}$ with the components of $N$-field which are the field-strength of
the potential $M_{m_1 \ldots m_6}$:

$$ N_{\alpha\beta a_1 \ldots a_5} =
 - (\Gamma_{a_1\ldots a_5})_{\alpha\beta}\,,\; N_{abc} = T_{abc} \eqno(A.2) $$
where
$$ N_{abc} \equiv {1\over 7!}\,{\epsilon_{abc}}^{b_1\ldots b_7}
 N_{b_1\ldots b_7}\,,       $$
all other components of $N$ are equal zero.

Equation of motion for the graviton ${e_m}^a$:

$$ \phi\,{\cal R}_{ab} + D_{(a}D_{b)}\,\phi -
{1\over 36}\,\phi\,\eta_{ab} T^2 +
 T_{c\,(a}\Gamma^c\Gamma_{b)}\,\chi  = 0         \eqno(A.3) $$

Equation of motion for the gravitino ${\psi_m}^\alpha$:

$$ \phi\,\Gamma^bT_{ab} - D_a\chi - {1\over 36}\,\Gamma_a{\hat T}\,\chi -
{1\over 24}\,{\hat T}\Gamma_a\,\chi = 0          \eqno(A.4) $$

Equation of motion for the $M_{m_1 \ldots m_6}$:

$$ D_{[a}(\,\phi\,T_{bcd]}) + {3\over 2}\,T_{[ab}\Gamma_{cd]}\,\chi +
{3\over 2}\,\phi\,T_{[abcd]}^2 = 0                 \eqno(A.5) $$

Together with the equations (3), (4)

$$ {\cal R} - {1\over 3}\,T^2 = 0              \eqno(A.6) $$

$$ \Gamma^{ab}T_{ab} = 0                      \eqno(A.7) $$
they form the complete system of equations of motion defining the fields
dynamics.

Although the $\phi$ and $\chi$ enter in the lagrangian in the noncanonical
manner, they also obey the wave equations which can be easy derived
from (A.3) + (A.6) + (A.7) and (A.4) + (A.7):

$$ D^a D_a\,\phi + {1\over 18}\,\phi\,T^2 = 0      \eqno(A.8) $$
$$ {\hat D}\,\chi + {1\over 9}\,{\hat T}\,\chi = 0 \eqno(A.9) $$

\bigskip

{\bf Appendix B: 0-components of superfields}

\bigskip

The method of transition from superfields to 0-components is
the standard one and has been described in \cite{9}. We present here only
the final result. The vertical line denotes a 0-component of
a superfield.

The supervielbein is defined as:

$$ {E_m}^a\vert = {e_m}^a,\;\;\;
{E_m}^\alpha\vert = {1\over 2}\,{\psi_m}^\alpha $$
$$ {E_\mu}^a\vert = 0,\;\;\;{E_\mu}^\alpha\vert =
\delta_\mu^\alpha                                  \eqno(B.1)  $$
The spin-connection ${\phi_{Ma}}^b$, corresponding to $D_M$, is
defined by
$$ D_M V^a = \partial_M V^a + V^b{\phi_{Mb}}^a\,,  \eqno(B.2)  $$
where $M=(m,\mu)$, $V^a$ -- is a vector. We suppose:

$$ {\phi_{ma}}^b\vert = {\omega_{ma}}^b\,,\;\;\;
{\phi_{\mu a}}^b\vert = 0                          \eqno(B.3)  $$
In the system of constraints (A.1) the spin-connection
$\omega_{abc}={e_a}^m\omega_{mbc}$ takes the form:

$$ \omega_{abc} = \omega_{abc}^{(0)} + {1\over 2}\,T_{abc}\vert +
C_{abc}\;,                                            \eqno(B.4)  $$
where
$$ C_{abc}={1\over 8}\,(\,2\,\psi_a\Gamma_{[b}\psi_{c]} +
\psi_b\Gamma_a\psi_c\,)\,,  $$
$\omega_{abc}^{(0)}$-- is the ordinary spin-connection depending
only on ordinary vielbein ${e_m}^a$.

The torsion ${T_{ab}}^\alpha$ in the system of constraints (A.1) is:

$$ T_{ab}\vert = \psi_{[b;\,a]} - {1\over 144}\,(\,\Gamma_{[a}{\hat T} +
3 {\hat T}\Gamma_{[a})\vert\psi_{b]} +
{1\over 4}\,\Gamma^{cd}\psi_{[a} C_{b]cd}\;,          \eqno(B.5) $$
where the semicolon denotes the ordinary covariant derivative with the
spin-connection $\omega_{abc}^{(0)}$ .

The relation between the $N$-field components with world and flat indices
 (see (A.2)) leads to:

$$ T_{abc}\vert = N_{abc}\vert = M_{abc} -
{1\over 8}\,\psi^d\Gamma_{dabcf}\psi^f              \eqno(B.6) $$
where
$$M_{abc} \equiv {1\over 6!}\,{\epsilon_{abc}}^{d_1\ldots d_7}
M_{d_1\ldots d_6;\,d_7}\;,\;\;M_{a_1\ldots a_6} \equiv
{{e_{a_1}}^{m_1}}\ldots{{e_{a_6}}^{m_6}}M_{m_1\ldots m_6}\;.   $$

The supercurvature ${{\cal R}_{mna}}^b$, corresponding to the
spin-connection ${\phi_{ma}}^b$, is:

$$ {{\cal R}_{mna}}^b = 2\,\partial_{[m}\,{\phi_{n]a}}^b -
2\,{\phi_{[m{\vert a \vert}}}^c\,{\phi_{n]c}}^b          \eqno(B.7)  $$
The explicit expression for the 0-component of ${\cal R}_{abcd}\vert$
 through the ordinary curvature $R_{abcd}$ , corresponding to the
spin-connection $\omega_{abc}^{(0)}$ , is very cumbersome. So we present
here only the expression for ${\cal R}\vert={{\cal R}_{ab}}^{ab}\vert$:

$$ {\cal R}\vert = R - {1\over 2}\,\psi_a\Gamma^{abc}\psi_{c;\,b} +
{1\over 2}\,(\psi^a\Gamma_b\psi^b)_{;\,a} +
{1\over 4}\,T_{abc}^2\vert + {1\over 8}\,\psi_a\Gamma_b\psi_c M^{abc} +  $$
$$ + {1\over 64}\,{(\psi_a\Gamma_b\psi_c)}^2 +
{1\over 32}\,(\psi_a\Gamma_b\psi_c)(\psi^a\Gamma^c\psi^b) -
{1\over 16}\,{(\psi_a\Gamma_b\psi^b)}^2                       \eqno(B.8) $$

\end{document}